# Photonic Bound-States-in-the-Continuum Observed with an Electron Nanoprobe

Zhaogang Dong[1,†,*], Zackaria Mahfoud[1,†], Ramón Paniagua-Domínguez[1], Hongtao Wang[2], Antonio I. Fernández-Domínguez[3], Sergey Gorelik,[1,4] Son Tung Ha[1], Febiana Tjiptoharsono[1], Arseniy I. Kuznetsov[1], Michel Bosman[1,5,*], and Joel K. W. Yang[1,2,*]

[1]Institute of Materials Research and Engineering, A*STAR (Agency for Science, Technology and Research), 2 Fusionopolis Way, #08-03 Innovis, 138634 Singapore

[2]Singapore University of Technology and Design, 8 Somapah Road, 487372, Singapore

[3]Departamento de Física Teórica de la Materia Condensada and Condensed Matter Physics Center, Universidad Autónoma de Madrid, 28049 Madrid, Spain

[4]Singapore Institute of Food and Biotechnology Innovation, A*STAR (Agency for Science, Technology and Research), 31 Biopolis Way, #01-02 Nanos, Singapore 138669

[5]Department of Materials Science and Engineering, National University of Singapore, 9 Engineering Drive 1, 117575, Singapore

*Correspondence and requests for materials should be addressed to J. K. W. Y. (email: joel_yang@sutd.edu.sg; telephone: +65 64994767), M. B. (email: msemb@nus.edu.sg; telephone: +65 65167563) and Z. D. (email: dongz@imre.a-star.edu.sg; telephone: +65 63194857).

**Abstract**

Bound-states-in-the-continuum (BIC) is a wave-mechanical concept that generates resonances with vanishing spectral linewidths. It has many practical applications in optics, such as narrow-band filters, mirror-less lasing, and nonlinear harmonic generation. As true BIC optical modes are non-radiative and confined to the near field of nanostructures, they cannot be excited using propagating light. As a result, their direct experimental observation has been elusive. Rather than using light, we demonstrate probing BIC modes on arrays of silicon nanoantennas using a focused beam of electrons in a transmission electron microscope. By combining cathodoluminescence (CL) and monochromated electron energy-loss spectroscopy (EELS) with controlled nanofabrication, we provide direct experimental evidence of "true" BIC modes, and demonstrate a BIC mode in the visible spectrum at 720 nm. The ability to observe and quantify these guided resonances with a spatial precision more than two orders of magnitude higher than previous far-field measurements allows the probing of individual elements in the nano-antenna arrays. The high-resolution experimental results are supported by numerical simulations as well as multipolar decomposition analysis, allowing us to demonstrate that the coherent interaction length of the quasi-BIC resonance requires at least 6 neighboring antenna elements, achieving over 60 times higher emissivity than for unpatterned silicon.

Bound-state-in-the-continuum (BIC) is a fascinating concept that has its origins in quantum mechanics in 1929.[1] Counterintuitively, optical modes with energies higher than the potential wells, *i.e.,* in the freely propagating continuum, can still be localized and bound in space when these wells are appropriately designed.[1] This BIC concept was first demonstrated experimentally in semiconductor heterostructures in 1992,[2] based on earlier theoretical predictions.[3, 4] To date, the concept of BIC has been generalized as a wave phenomenon into various fields, such as acoustics,[5, 6] microwave physics,[7, 8] and optics.[9-13] The BIC concept in optics exhibits itself in remarkably narrow resonances that are controlled by the design of nanostructures, such as dielectric gratings,[9] arrays of coupled waveguides,[10, 11] layered nanoparticles,[12] and photonic crystal slabs.[13] The general principle of BIC is the complete suppression of radiative losses in the far field by total destructive interference of radiation from the charge-current configuration associated to the mode.[14, 15] These BIC modes have energies above the light line, hence the "continuum" denomination, but are completely "dark" in the sense that they do not couple to radiative modes in free-space but are instead bound to the structures that support them.

Recently, the BIC concept has been extensively explored in nanophotonics mostly with nanoantenna arrays,[19-25] but uniquely in individual resonators too.[12, 16-18] Nanoantenna designs with broken symmetry provide small but sufficient coupling to form quasi-BIC modes that enable far-field optical excitation, while true BIC modes do not couple to the far field.[20-26] Due to the (quasi-) BIC modes being supported by nanoantennas, it is possible to achieve unidirectional scattering,[27] sensitive hyperspectral imaging,[24] lasing,[28-30] nonlinear nanophotonics,[17, 23, 31, 32] chirality,[33] bio-sensing,[24, 25, 34] and topological photonics.[26, 27] However, due to the "dark" nature of the true BIC resonance, optical excitation in the far field is unable to reveal its mode characteristics. In addition, although it has been demonstrated that a finite antenna array with only 8 × 8 elements is able to achieve BIC lasing under far field optical excitation,[35] no direct characterization technique has determined the characteristic length that is required to set-up a BIC mode, or rather how quickly the BIC mode decays as it approaches the edge of a large array.

BIC modes have no coupling to free-space radiation and are therefore optically inaccessible, except for the quasi-BIC modes. Experiments that attempt to measure BIC modes will therefore need to be sensitive to both the near-field and the far-field simultaneously; in this way, the "true", non-radiative BIC modes can be distinguished from the quasi, radiative BIC modes. To this aim, our experiments combine cathodoluminescence (CL) and monochromated electron energy loss spectroscopy (EELS) in a scanning transmission electron microscope (STEM). The nanometer-sized, focused electron beam is used to excite the optical modes in the near-field. The energy transferred in these radiative and non-radiative excitations is measured with EELS, while only radiative losses are measured with CL in the far-field. A comparison of both EELS and CL spectra distinguishes the lossy and the trapped optical modes, providing an experimental set-up to unambiguously characterize true photonic BICs with nanometer spatial precision.

In this work, we provide experimental and theoretical evidence of true BIC resonances that are locally excited with a nanometer electron beam in the STEM and directly probed with EELS and CL spectroscopy. Localized plasmonic optical transitions have been investigated before in the STEM.[36-39] The electron microscope was shown to provide a unique combination of broadband, near-field excitation, wide-range spectroscopy, and ultra-high spatial resolution, enabling a comprehensive analysis of optical resonances. Here, we use it to probe the BIC mode on designed arrays of silicon nanoantennas. By combining the spectroscopic techniques CL and EELS in the STEM, we provide direct experimental evidence of "true" BIC modes. We place "true" in quotation marks, as our fabricated arrays are necessarily of finite size, while true BIC modes in the strict term are only fully-developed in infinitely-large arrays. Numerical simulations as well as multipolar decomposition analysis support our claim that quasi- and "true" BIC modes can be distinguished. Our approach is extendable beyond BIC, and is applicable to other emerging optical phenomena such as the localized probing of quantum emitters and other engineered nanostructures.

**Localized Excitation and Probing of BIC Modes**

Fig. 1a shows the experimental setup for probing the BIC mode, where a high-energy electron beam with a diameter of ~1 nm in the STEM is focused onto an array of Si nanoantennas, lithographically defined on a 30-nm-thick suspended amorphous $Si_3N_4$ membrane (see Supplementary Fig. 1 and Methods section for fabrication details). When this focused electron beam is positioned near a nanostructure, a fraction of the energy of the electron beam will be used to polarize the free and bound electrons within the material that start to oscillate according to the modes supported by the nanostructures. These resonant modes can subsequently lose their energy to free-space emission, which is measured with CL,[40-43] as shown previously for metallic nanostructures[43-49] and silicon nanoantennas.[50-54] All the losses—radiative and non-radiative—that the electron beam encounters in the excitation of the resonant modes are measured experimentally with EELS, a true near-field technique. By combining both spectroscopy techniques in the STEM, we will demonstrate the possibility of measuring BIC modes at the nanometer length scale. Our universal approach to distinguish all the bright and dark optical modes has the additional advantage of performing simultaneous STEM imaging with nanometer resolution, to visualize the local sample morphology. A more detailed description of the experimental set-up for measuring "true" and quasi-BIC modes is given in the Methods.

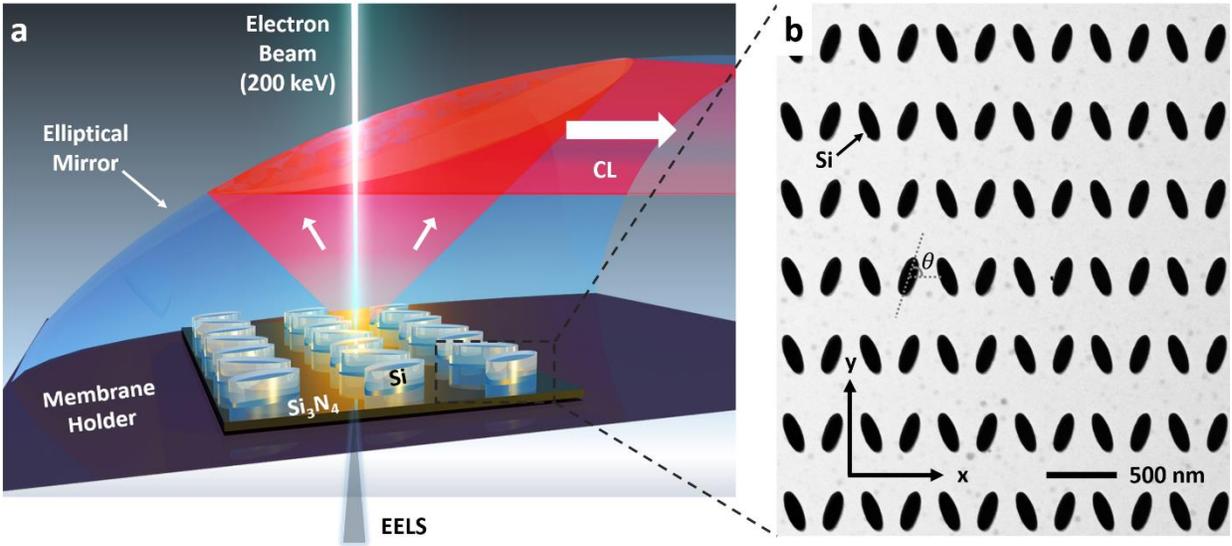

**Fig. 1 | Experimental setup of electron energy loss spectroscopy (EELS) and cathodoluminescence (CL) measurements for probing the Si nanoantenna arrays supporting bound-states-in-the-continuum (BIC). a**, Schematic of the experimental setup in an STEM where a high-energy electron beam focused to ~1 nm is used to probe an array of Si nanoantennas on a 30-nm-thick suspended $Si_3N_4$ membrane. Complementary measurements of the energy lost by electrons, and energy of emitted photons result in EELS and CL spectra respectively, though not collected simultaneously. **b**, Bright-field STEM image of the amorphous Si nanoantenna array supporting quasi-BIC modes on a 30-nm-thick $Si_3N_4$ membrane. "True" BIC occurs when $\theta = 90°$.

Fig. 1b shows an STEM image of an array of Si elliptic cylinder nanoantenna supporting (quasi-)BIC modes. The designed dimensions for the long-axis and short-axis of the nanoellipses are 286 nm and 96 nm respectively, with a pitch size of 562 nm. The tilt angle θ of the nano antennas is commonly used to control the degree of symmetric breaking of the coupled nanoantennas,[22] providing far-field optical access to the quasi BIC mode. As θ approaches 90º (see the STEM image in Fig. 2a), the silicon nanoantenna array supports a "true" BIC mode that is not optically accessible from the far field. The near-field pattern of this "true" BIC mode can be found by computing the eigenmodes of the system. Fig. 2b shows the simulation results, corresponding to a BIC mode at the Gamma symmetry point at a wavelength of ~720 nm and a quality factor of ~90. Note that the finite bandwidth of the mode is due to the dissipative loses of Si at that wavelength. The near-field oscillation characteristic of this BIC mode is shown in the supporting video file "NF_BIC.gif", where the neighboring unit cells are oscillating in phase when $k$=0. The schematic in Fig. 2c shows the equivalent multipole moments that give rise to the BIC, consisting of a $z$-polarized magnetic dipole ($MD_z$) and an electrical quadrupole ($EQ_{xy}$). Due to its non-radiative nature, this "true" BIC resonance mode can neither be excited nor measured using far-field techniques.[16] In our experimental set-up, the electron energy losses measured with monochromated EELS would detect this BIC mode in the near-field. On the other hand, in-situ CL measurements that collect emitted light will allow us to determine if this mode can be optically detected in the far field.

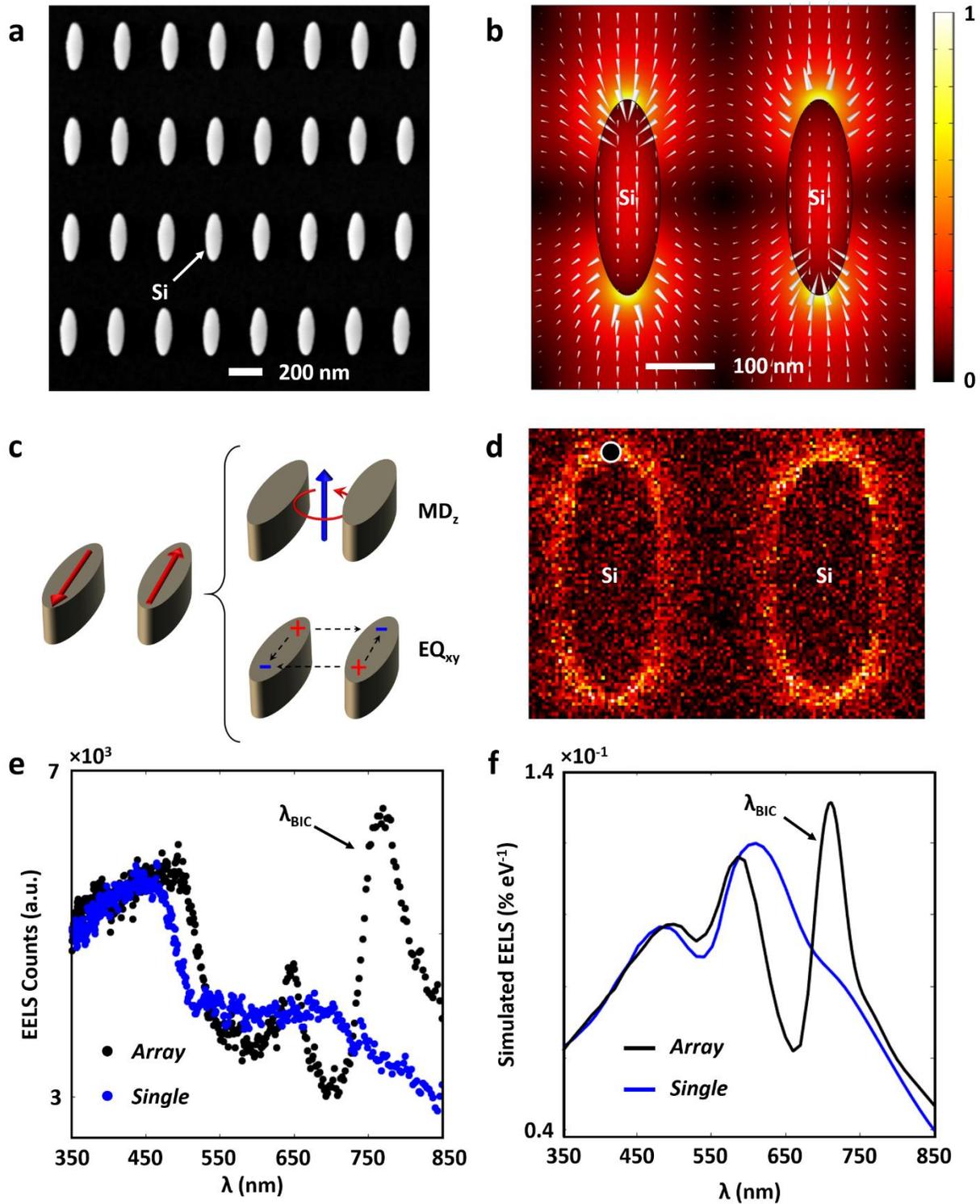

**Fig. 2 | Probing of a "true" BIC resonance mode by Electron Energy Loss Spectroscopy. a**, STEM annular dark field image of the silicon nanoantenna array with elliptic cylinders aligned at

a tilt angle θ=90°. **b**, Simulated near-field mode pattern of the true BIC mode, with arrows indicating the electric field direction; their length being proportional to the amplitude. **c**, Schematic of the multipole moments comprising the "true" BIC mode: the $z$-polarized magnetic dipole ($MD_z$) and the electric quadrupole ($EQ_{xy}$). **d**, Experimental EELS map of the elliptic cylinder nanoantenna array with the "true" BIC mode resembling the E-field maps in (b). **e**, EELS spectrum (black color) as measured at the position near the tip of the ellipse as indicated in (d) demonstrating the excitation of the "true" BIC mode. For benchmarking, the EELS spectrum measured from an isolated single antenna element of the same size (blue color) clearly shows the missing peak associated with the BIC mode. **f**, FEM-simulated EELS spectra for the nano antenna array (black), showing a BIC resonance, and for a single antenna element (blue), showing no BIC resonance.

EELS measurements of a "true" BIC mode are presented in Fig. 2. The measured EELS BIC map in Fig. 2d and EELS spectra in Fig. 2e show clear evidence of a BIC mode in the silicon antenna array, while this mode is strikingly absent when EELS is measured on individual antennas that are not part of an array, confirming the interpretation of the array effect. To corroborate these experimental observations, Fig. 2f presents finite-element-method (FEM) -based COMSOL simulations of EELS spectra for a single antenna and for antennas in an array (see Methods). The simulation confirms the observed difference between the antenna array and the single element. Moreover, the first row in Fig. 3b presents optical reflectance measurements under $x$-polarized incidence condition (see the $x$-$y$ coordinate in the STEM image in Fig. 1b), where no BIC resonance feature is observed at ~720 nm in the case of a 90° tilt angle due to its non-radiative nature. Final evidence is provided in Fig. 3c where no BIC resonance is observed as CL far field emission, even though the EELS spectra in Fig. 3a show that the mode is excited by the electron beam. This is direct verification of the non-radiative nature of this "true" BIC mode.

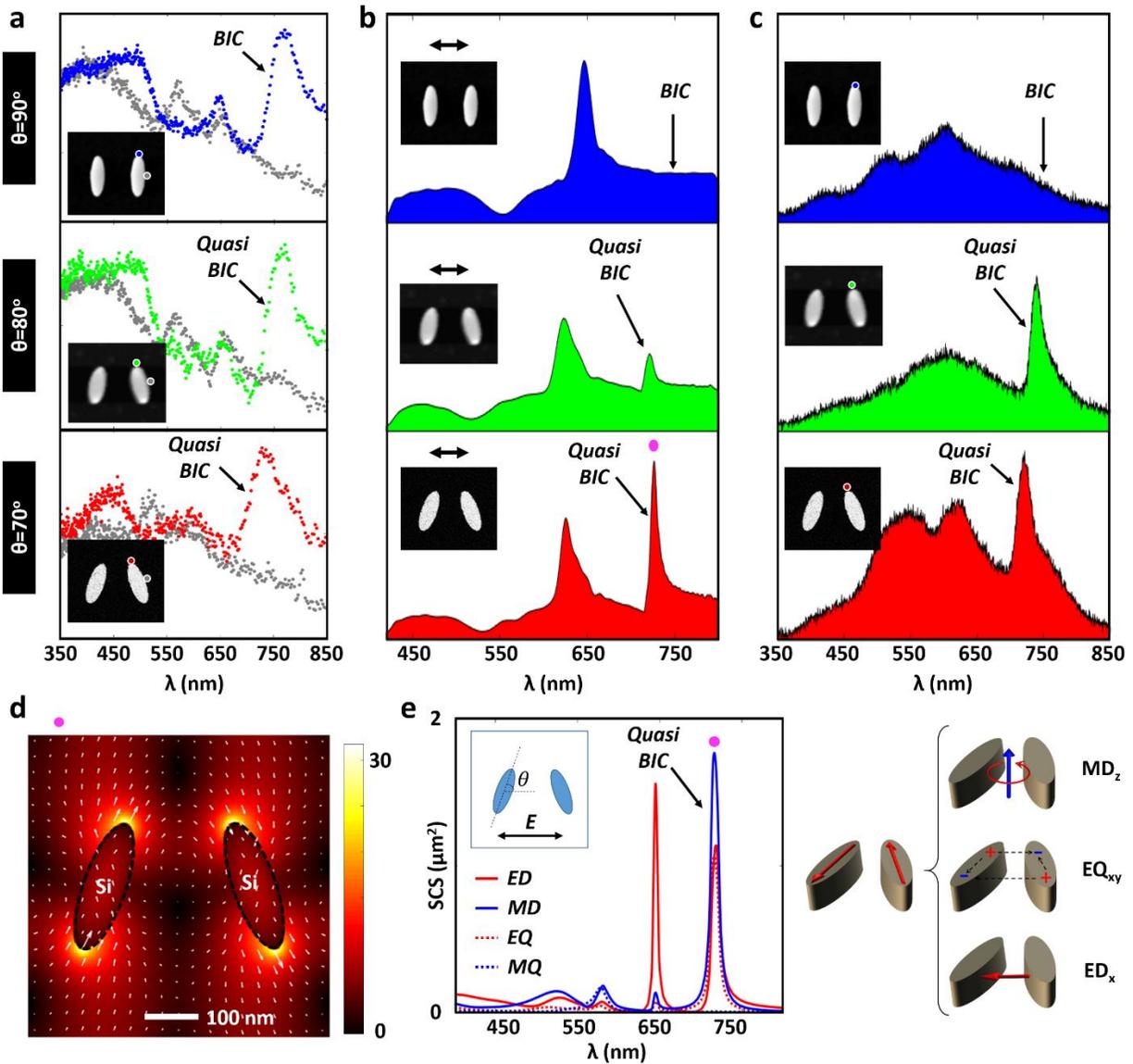

**Fig. 3 | Evolution of CL and EELS emission spectra for the nanoantenna array with different tilt angles. a**, Measured EELS spectra, **b**, reflectance spectra under *x*-polarized incidence condition, and **c**, CL spectra from the nanoantenna array with different tilt angle θ of 90°, 80°, and 70°. **d**, Spatial distribution of the electric field |***E***| at the quasi BIC resonance at ~720 nm for the nanoantenna array with a tilt angle θ of 70°. The arrows denote the electric field direction, the length of the arrows being proportional to the amplitude. **e**, Multipolar decomposition of the

scattering cross-section (SCS) and corresponding schematics of the multipoles excited at the quasi-BIC mode.

In an attempt to distinguish the "true" bound states from the emissive quasi-BIC resonances, we patterned symmetry-broken arrays of the same nanoantennas by changing the tilt angle θ of the nano antennas from 90º, to 80º and 70º. Fig. 3a presents the measured EELS spectra showing both the "true" BIC resonance at θ=90º and the quasi-BIC resonances at θ=80º and θ=70º. The quasi-BIC mode now becomes visible both in the reflectance spectra and in the CL emission spectra, as shown in Figs. 3b and 3c, while this mode is not present for the "true" BIC at θ=90º. For completeness, the simulated reflectance spectra are shown in Supplementary Fig. 2. For the case when θ=70º, the quasi-BIC mode is particularly strong in CL emission, and has a full-width-at-half-maximum (FWHM) of ~28 nm (see Supplementary Fig. 3 for details). The local field distribution for this quasi-BIC resonance is given in Fig. 3d, showing a ~23-fold local electric field enhancement factor. Fig. 3e presents the corresponding multipolar decomposition, showing how electric dipole (*ED*), magnetic dipole (*MD*), electric quadrupole (*EQ*) and magnetic quadrupole (*MQ*) contributions are excited. The magnetic dipole is directed along the *z*-direction ($MD_z$), while the electric dipole lies in-plane along the *x*-direction ($ED_x$), allowing the in- and out-coupling of the incident plane-wave radiation. Therefore, the nano-optical setup of STEM-EELS in combination with CL is able to differentiate and characterize both non-radiative "true" BIC resonances as well as the radiative quasi-BIC resonances.

**Coherent Interaction Length of the Quasi-BIC Resonance**

One important feature of the quasi BIC resonance in nanoantenna arrays is that it is a collective array mode, whose formation involves a number of neighboring nanoantenna elements.

To show this array effect, we investigate the coherent interaction length of the quasi-BIC resonance based on the CL characterization results in a Si nanoantenna array. No EELS mapping was used here, as the monochromated EELS signal is much weaker and thus requires extremely long data acquisition times. The CL mapping was carried out on a Si nanoantenna array containing a 130×130 antenna elements, the corner of which is shown in the STEM high-angle annular dark-field (HAADF) image in Fig. 4a, where the Si nanoantenna array has a tilt angle θ of 70º.

Fig. 4b was obtained by scanning the electron beam in the white dotted rectangle of Fig. 4a and recording a CL spectrum at each pixel in the scan. This three-dimensional CL data set is then stored and processed as described in the Methods section, giving in Fig. 4b the spatial distribution of the CL emission at the quasi-BIC resonant wavelength of ~722 nm, using an integration bandwidth of 17 nm. Fig. 4c presents the CL spectra measured from a single antenna element indicated in the white dashed rectangle in Fig. 4b. It shows that the CL emission measured at the tip position of the silicon nanoantenna (in red color) is significantly different from the CL spectrum measured at the middle position (in black color); these respective measurement positions are indicated on the inset STEM image in Fig. 4c. For comparison, the graph also plots the CL emission measured from a 90-nm-thick, *unpatterned* Si film; its much lower CL intensity is measured to give a 63-times weaker emissivity as compared to the Si array.

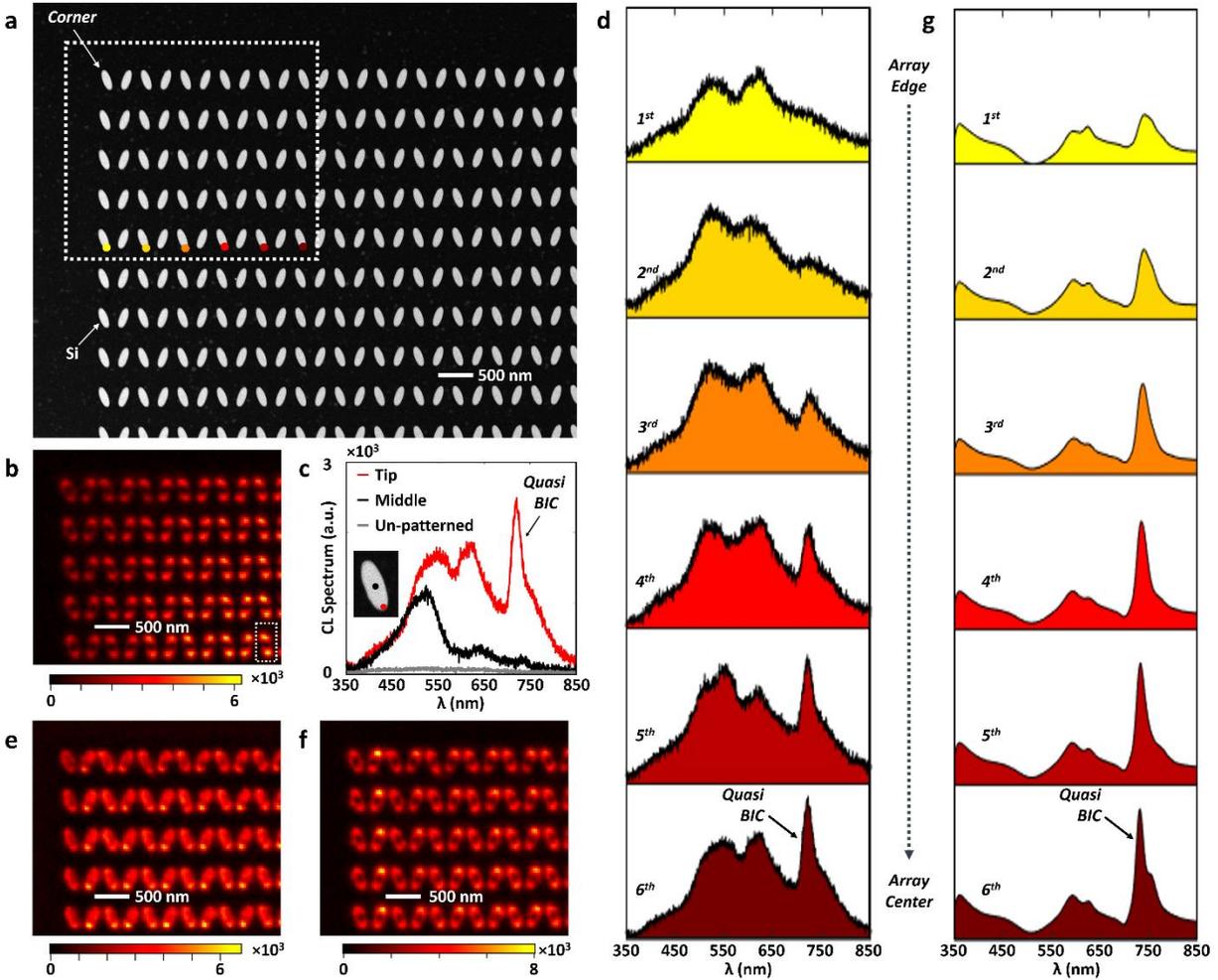

**Fig. 4 | CL characterization of the coherent interaction length of the quasi-BIC resonance in a Si nanoantenna array with a tilt angle θ of 70°. a**, STEM HAADF image, showing the corner of a fabricated Si nanoantenna array with 130×130 elements. The white dotted rectangle highlights the area for CL mapping experiment and the color dots denote the measurement positions for investigating CL spectrum evolution. **b**, CL intensity map at 722 nm, the quasi-BIC resonance wavelength, by scanning the focused electron beam over the Si nanoantenna array. The wavelength bandwidth for the CL integration is 17 nm. **c**, CL spectra from the "tip" position and "middle" position of one selected antenna as highlighted by the white dashed line in (b). The 722 nm emission at the "tip" position is strongly enhanced due to the quasi-BIC resonance. The CL

spectrum as measured from un-patterned 90-nm-thick Si film is plotted in grey for benchmarking. **d**, Evolution of the measured CL emission spectra when going from the "array edge" towards the "array center". It shows that the quasi-BIC resonance requires the presence of neighboring array antennas. **e-f**, CL maps at the Mie resonance wavelengths of the individual antennas (e) 522 nm and (f) 614 nm. **g**, Simulation of the CL spectrum evolution.

Furthermore, the CL mapping at Fig. 4b gives direct visual evidence of the coherent interaction length of the quasi-BIC resonance due to collective antenna array effect. As seen there, the CL emission at the quasi-BIC (~722 nm wavelength) is very weak at the "corner" position of the antenna array, and it becomes stronger when the electron beam moves towards its center. This edge effect becomes even more obvious when we plot the CL spectra as a function of the distance to the edge of the array, shown in Fig. 4d. The 1$^{st}$ antenna element at the array edge shows no sign of the quasi-BIC resonance at all. Our CL experiments demonstrate that the quasi-BIC mode is only fully established at least six antenna elements into the array; the outer six elements only exhibit a partially-established quasi-BIC mode. This observation is in stark contrast with the CL map at the Mie resonances of the nanoantennas shown in Figs. 4e-f. As these resonances are supported by individual antennas, we observe no emission reduction or enhancement as a function of location in the nanoantenna array.

To corroborate this understanding, Supplementary Fig. 4 presents simulations of the CL emission as enhanced by the quasi BIC resonance. We adopted the approach from Das et al.,[55] performing FDTD simulations with the incident electron beam modeled as an array of *z*-polarized dipole sources with an appropriate phase delay (see details in the Methods section). Here, to account for the spectral broadening effect due to the limited CL spectral resolution, Fig. 4g was obtained after the convolution with a normalized Gaussian function (see details in Methods section). These simulation results in Supplementary Fig. 4 and Fig. 4g verify the experimental observations in Figs. 4a-d, showing that the first six antenna elements at the edge of the array only have partially-established quasi-BIC resonances. In addition, the EELS calculations for nanoparticle arrays of different sizes shown in Supplementary Fig. S5 indicate that the interaction coherent length for "true" BIC is similar to the one extracted from CL measurements and calculations. This indicates that by tilting the ellipse orientation, the modes acquire certain radiative character, but they remain very similar to their fully dark counterpart for $\theta=90°$.

**Conclusions**

By combining nanoscale electron beam spectroscopy in the STEM, theoretical simulations and multipolar decomposition, we observe "true" photonic bound-states-in-the-continuum on nanostructured Si samples. This localized probing approach provides several distinguishing advantages over conventional optical excitation approaches in the far field. For instance, the electron beam-induced CL emission is able to spatially map the coherent interaction length of the quasi-BIC resonance, which, in our case, shows that this mode is only partially-established in the outer six antenna elements, and only fully formed beyond the sixth element towards the center of the array. When STEM-CL spectroscopy is combined with monochromated STEM-EELS measurements, the "true" BIC resonance can be revealed, even in situations when their excitation is impossible using far-field optical irradiation. Our results provide a general methodology to quantitatively probe the mode formation mechanism for BIC with nanometer spatial precision. It will be broadly applicable to other emerging optical resonances and will provide direct and unique insight into their nature, and may even be used to enable and quantify the localized excitation of quantum emitters.

**Methods**

**Nanofabrication of Si Nanostructured on Membrane.** Amorphous Si with a thickness of 90 nm was grown onto a 30-nm-thick $Si_3N_4$ membrane (Agar Scientific S1711), by using plasma-enhanced chemical vapor deposition (PECVD) method. Before the electron beam exposure, HSQ resist with a concentration of 2% wt., diluted in methyl isobutyl ketone (Dow Corning XR-1541-002), was spin coated onto the cleaned substrate at 5k round-per-minute (rpm) giving a thickness of ~30 nm. The electron beam exposure was carried out with the following conditions: electron acceleration voltage of 100 keV, beam current of 200 pA, and the exposure dose of ~12 mC/cm$^2$. The sample was then developed by a NaOH/NaCl salty solution (1% wt./4% wt. in de-ionized water) for 60 seconds and immersed in de-ionized water for 60 seconds to stop the development. Next, the sample was immediately rinsed by acetone, isopropanol alcohol (IPA) and dried by a continuous flow of nitrogen. Si etching was then carried out by using inductively-coupled-plasma (ICP, Oxford Instruments Plasmalab System 100), with $Cl_2$ gas chemistry at 6 °C.[56]

**Cathodoluminescence (CL) and electron energy loss spectroscopy (EELS) in a scanning transmission electron microscope (STEM).** The STEM-CL spectroscopy and STEM-HAADF imaging experiments were carried out with a FEI Titan with the sample at room temperature. The system is equipped with a Schottky emitter and a Wien-type monochromator that was used to disperse and filter the electron beam to a resolution of 65 meV. EELS spectra were acquired with a Tridiem HR detector in spectrum imaging mode, using binned gain averaging.[57] The background

signal was subtracted by fitting a high-quality 'zero-loss peak' to the measured spectra between 0.5 and 1.0 eV, well before the first onset of the experimental peaks. An example of the EELS background fitting is shown in Supplementary Fig. 6. A Gatan Vulcan (sample holder & light detection system) was used for CL measurements. The electron beam was accelerated at 200 keV with the monochromator being turned off, where the electron beam was focused into a probe of approximately 1 nm. The CL maps were acquired with a diffraction grating blazed at 500 nm and a dwell time set to 0.42 second per spectrum, where the CL spectral resolution is 17 nm, limited by the slit aperture of the spectrometer. Data processing consisted of subtracting the CCD dark-noise read-out and removing strong single-channel 'X-ray' spikes.

**Reflectance Simulations and Multipolar Decomposition.** Finite-difference time-domain (FDTD) simulations were carried out using a commercial software (Lumerical FDTD Solutions). In order to consider the random deformation of the ultra-thin $Si_3N_4$ membrane (*i.e.* 30 nm in thickness) during the ICP CVD process for growing Si film, 13×13 arrays of Si nanoantennas were used with the incident optical field being linearly polarized along *x*-direction at normal incidence condition. A non-uniform meshing technique was applied with a minimum mesh size down to 0.5 nm. The dielectric function of amorphous Si was taken from the measured *n* and *k* values by ellipsometry.[58] Moreover, multipolar decomposition analysis was carried out to analyze the optical modes being excited within the structures, where this multipolar decomposition was implemented by Finite Element Method (FEM) in COMSOL. The detailed formulas of multipolar analysis can be found in the supporting information of the reference.[58]

**Cathodoluminescence (CL) Simulations.** The simulation of CL emission was based on a finite-difference time-domain (FDTD) approach as reported in the literature,[55] where an array of dipole sources is able to simulate the CL emission process. Supplementary Fig. 4a presents the schematics, using an array of 41 $z$-polarized dipole sources distributed with a uniform separation of 50 nm along the $z$ direction and having relative phase-shift following the expression $e^{i\omega z/v}$. Here, $v$ denotes the electron speed, $\omega$ denotes the oscillation frequency and $z$ denotes the dipole position.[55] These 41 $z$-polarized dipole sources are placed symmetrically with respect to the 30-nm-thickness membrane. At our TEM setup, the acceleration electron voltage is 200 kV, and the electrons are travelling at a speed of ~0.695c. To record the CL emission, a monitor is placed at a $z$-plane, which is 500 nm above the top surface of the 13×13 elliptic cylinder nanoarray. In order to remove the direct emission of these dipole sources, we take the vectorial subtraction between the total electrical field components with and without the elliptic cylinder nanoarray. After that, a near-to-far transformation was carried out to simulate the collected CL emission within a solid angle of (-15°, 15°). The detailed CL simulation results are shown in Fig. S4. In addition, to account for the spectral broadening effect due to the limited spectral resolution for CL measurements, this simulated CL spectrum were then convoluted with a normalized Gaussian function of 16 meV width to get the simulated CL emission profiles as shown in Fig. 4g.

**Electron Energy Loss Spectroscopy (EELS) Simulation.** The numerical calculations were performed under the finite-element solver of Maxwell's Equations in frequency domain implemented in COMSOL Multiphysics. The frequency-dependent electron beam propagating along $z$-direction at position $(x_e, y_e)$ was simulated by a line current of the form $\boldsymbol{j}(\boldsymbol{r},\omega) = e \exp\left\{-\frac{i\omega z}{v}\right\} \delta(x - x_e)\delta(y - y_e)\hat{\boldsymbol{z}}$, where $e$ is the electron charge and the electron velocity, $v$, was set in accordance with the experiments. The energy loss probability is then calculated as $A = \Gamma = \frac{1}{\pi\hbar\omega} \int dz\, Re\, \{j^*(r,\omega) E_{SC}(r,\omega)\}$, where $E_{sc}(r, \omega)$ are the electric fields scattered by the dielectric nanoparticle array.[40] The EELS spectra were calculated by convolving $\Gamma(\omega)$ with a normalized Gaussian function of 65 meV width, mimicking the experimental spectral resolution. The detailed simulated EELS spectra for different array size are shown in Supplementary Fig. 5 for the case θ=90°.

**Data availability**

The data that support the figures and other findings of this study are available from the corresponding authors upon reasonable request.

**Acknowledgements**

We would like to acknowledge the funding support from Agency for Science, Technology and Research (A*STAR) SERC Pharos project (grant number 1527300025). In addition, Z.D. and J.K.W.Y. would like to acknowledge the A*STAR AME IRG funding support with the project number of A20E5c0093. Z.D. acknowledges the support from A*STAR career development award (CDA). M.B. acknowledges support from the Singapore Ministry of Education Academic Research Fund Tier 2 (project number MOE2019-T2-1-179). A.I.F.D. was supported by a 2019 Leonardo Grant for Researchers and Cultural Creators, BBVA Foundation. Z.M. acknowledges the support from the European Union's Horizon 2020 research and innovation program under grant agreement No 823717 – ESTEEM3.

**Author contributions**

Z.D. and J.K.W.Y. conceived the concept, designed the experiments and wrote the manuscript. Z.M. and M.B. performed the cathodoluminescence (CL) and electron energy loss spectroscopy (EELS) measurements, data processing, STEM and TEM imaging. R.P.-D.

performed the COMSOL simulations on the multipolar decomposition analysis and provided theoretical support of the eigenmode calculation for the true BIC resonance at $\theta=90^\circ$. H.W. performed the finite-difference time-domain (FDTD) simulations, including reflectance spectra simulation, CL spectrum simulations and optical mode patterns calculation. A.F.-D. performed the EELS simulation. S.G. performed the reflectance spectrum measurements. S.T.H. provided insightful suggestions to improve the experiments. F.T. and Z.D. performed the chemical vapor deposition (CVD) growth of amorphous Si onto the 30-nm-thick $Si_3N_4$ membranes and inductively coupled plasma (ICP) etching of Si based on $Cl_2$ gas chemistry. A.I.K. participated in discussions and gave suggestions. All authors analyzed the data, read and corrected the manuscript before the submission. Z.D and Z.M. contributed equally to this work.

**Competing interests**

The authors declare no competing interests.

**Additional information**

**Correspondence and requests for materials** should be addressed to J.K.W.Y., M.B. or Z.D.

# Supplementary Information for:

# Photonic Bound-States-in-the-Continuum Observed with an Electron Nanoprobe

Zhaogang Dong[1,†,*], Zackaria Mahfoud[1,†], Ramón Paniagua-Domínguez[1], Hongtao Wang[2], Antonio I. Fernández-Domínguez[3], Sergey Gorelik,[1,4] Son Tung Ha[1], Febiana Tjiptoharsono[1], Arseniy I. Kuznetsov[1], Michel Bosman[1,5,*], and Joel K. W. Yang[1,2,*]

[1]Institute of Materials Research and Engineering, A*STAR (Agency for Science, Technology and Research), 2 Fusionopolis Way, #08-03 Innovis, 138634 Singapore

[2]Singapore University of Technology and Design, 8 Somapah Road, 487372, Singapore

[3]Departamento de Física Teórica de la Materia Condensada and Condensed Matter Physics Center, Universidad Autónoma de Madrid, 28049 Madrid, Spain

[4]Singapore Institute of Food and Biotechnology Innovation, A*STAR (Agency for Science, Technology and Research), 31 Biopolis Way, #01-02 Nanos, Singapore 138669

[5]Department of Materials Science and Engineering, National University of Singapore, 9 Engineering Drive 1, 117575, Singapore

*Correspondence and requests for materials should be addressed to J. K. W. Y. (email: joel_yang@sutd.edu.sg; telephone: +65 64994767), M. B. (email: msemb@nus.edu.sg; telephone: +65 65167563) and Z. D. (email: dongz@imre.a-star.edu.sg; telephone: +65 63194857).

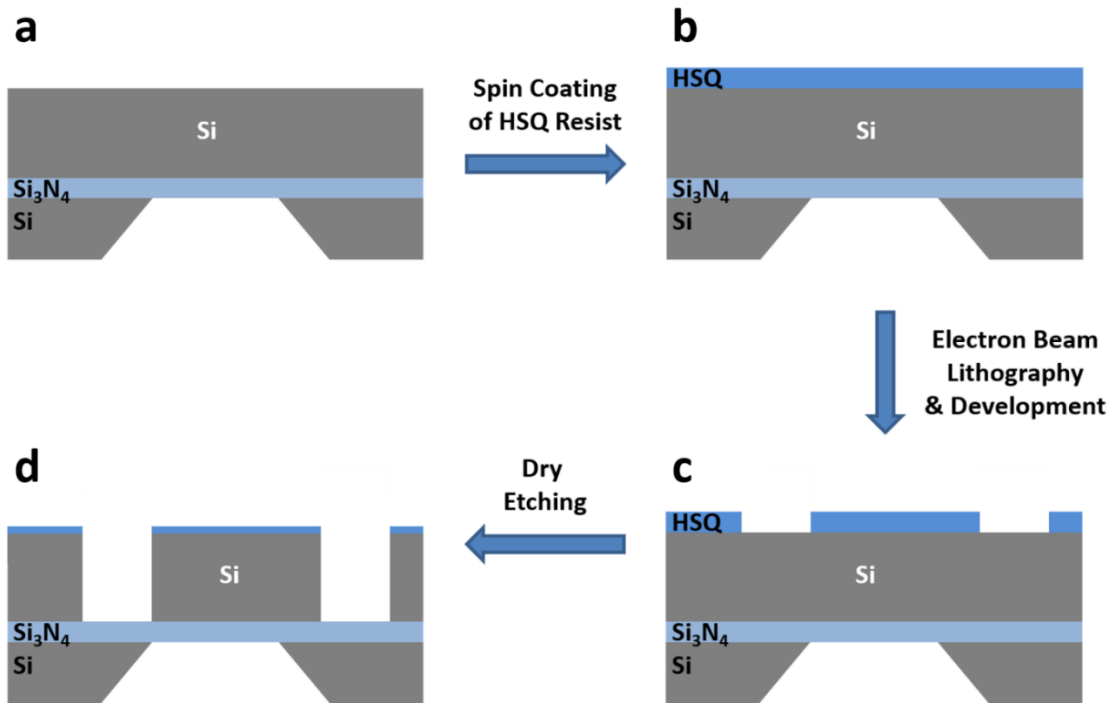

**Supplementary Fig. 1 | Fabrication process of amorphous silicon nanostructures on a 30-nm-thick Si$_3$N$_4$ membrane. a**, 90-nm-thick amorphous silicon film is grown onto the 30-nm-thick amorphous Si$_3$N$_4$ membrane by using plasma-enhanced chemical vapor decomposition (PECVD). **b**, 30-nm-thick hydrogen silsesquioxane (HSQ, 2%, Dow Corning XR-1541-002) resist is spin coated onto the amorphous silicon film surface at a speed of 5k rpm. **c**, E-beam exposure to fabricate the HSQ resist mask. **d**, Inductively-coupled plasma (ICP) for silicon etching with Cl$_2$ gas chemistry.

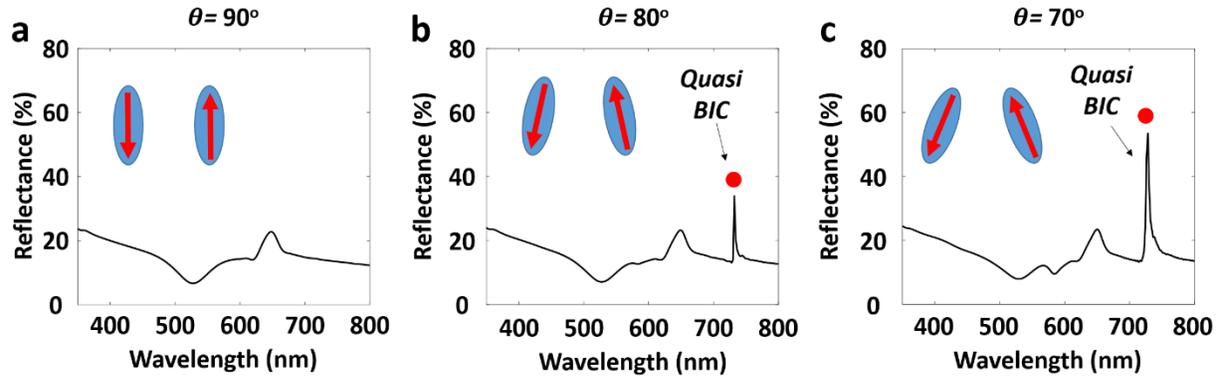

**Supplementary Fig. 2 | Simulated reflectance spectra of a pair of nanoellipses with different tilt angle θ.** The excitation optical field is *x*-polarized. **a**, θ=90º. **b**, θ=80º. **c**, θ=70º. It shows that the "true" BIC resonance cannot be excited when the angle θ is 90º. The inset figures schematically present the effective dipole moments induced in the silicon array elements at the "true" BIC and quasi-BIC condition.

**Supplementary Note 1. Fitting of CL emission profiles with Lorentz oscillator model.**

For the CL emission profile as shown in Fig. 3(c) for θ=70°, it could be fitted with three Lorentzian profiles with the following equations:

$$CL = C_1 \times \frac{\omega_p^2 \gamma_1 \omega}{(\omega_1^2-\omega^2)^2+\omega^2\gamma_1^2} + C_2 \times \frac{\omega_p^2 \gamma_2 \omega}{(\omega_2^2-\omega^2)^2+\omega^2\gamma_2^2} + C_3 \times \frac{\omega_p^2 \gamma_3 \omega}{(\omega_3^2-\omega^2)^2+\omega^2\gamma_3^2}, \quad (S1)$$

where $\omega$ represents the frequency. $\omega_1$, $\omega_2$ and $\omega_3$ are the resonant frequencies of the three Lorentz oscillator models. $\gamma_1$, $\gamma_2$ and $\gamma_3$ are to denote the respective damping coefficients, and $\omega_p$ represents the plasma frequency of silicon. The fitted values of $\omega_1$, $\omega_2$ and $\omega_3$ are 1.718 eV, 1.987 eV and 2.34 eV. In addition, the fitted values of $\gamma_1$, $\gamma_2$ and $\gamma_3$ are 0.067 eV, 0.23 eV and 0.59 eV. Therefore, the CL emission profile due to the quasi BIC mode at 1.718 eV (*i.e.* ~722 nm) has a full-width-at-half-maximum (FWHM) of ~28 nm.

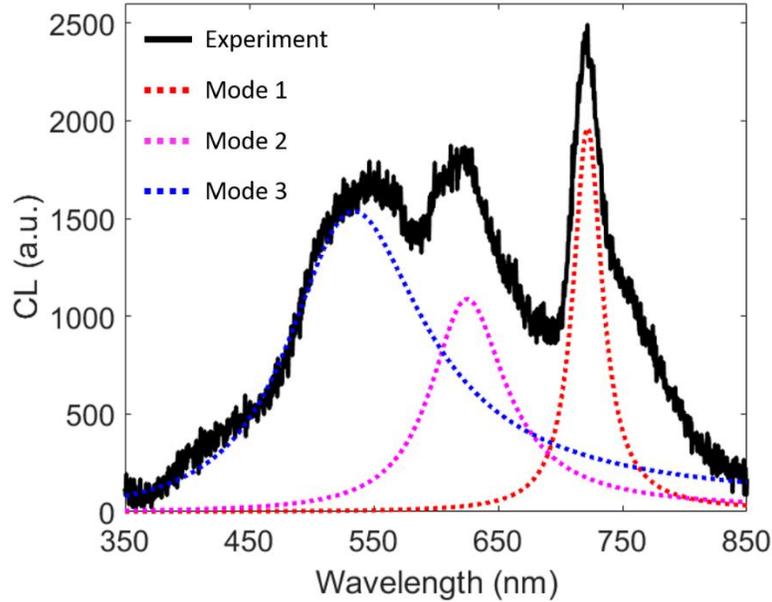

**Supplementary Fig. 3 | Fitting of the measured CL emission profile with three Lorentz oscillator models.** The CL emission profile due to the quasi BIC mode at 1.718 eV (*i.e.* ~722 nm) has a full-width-at-half-maximum (FWHM) of ~28 nm.

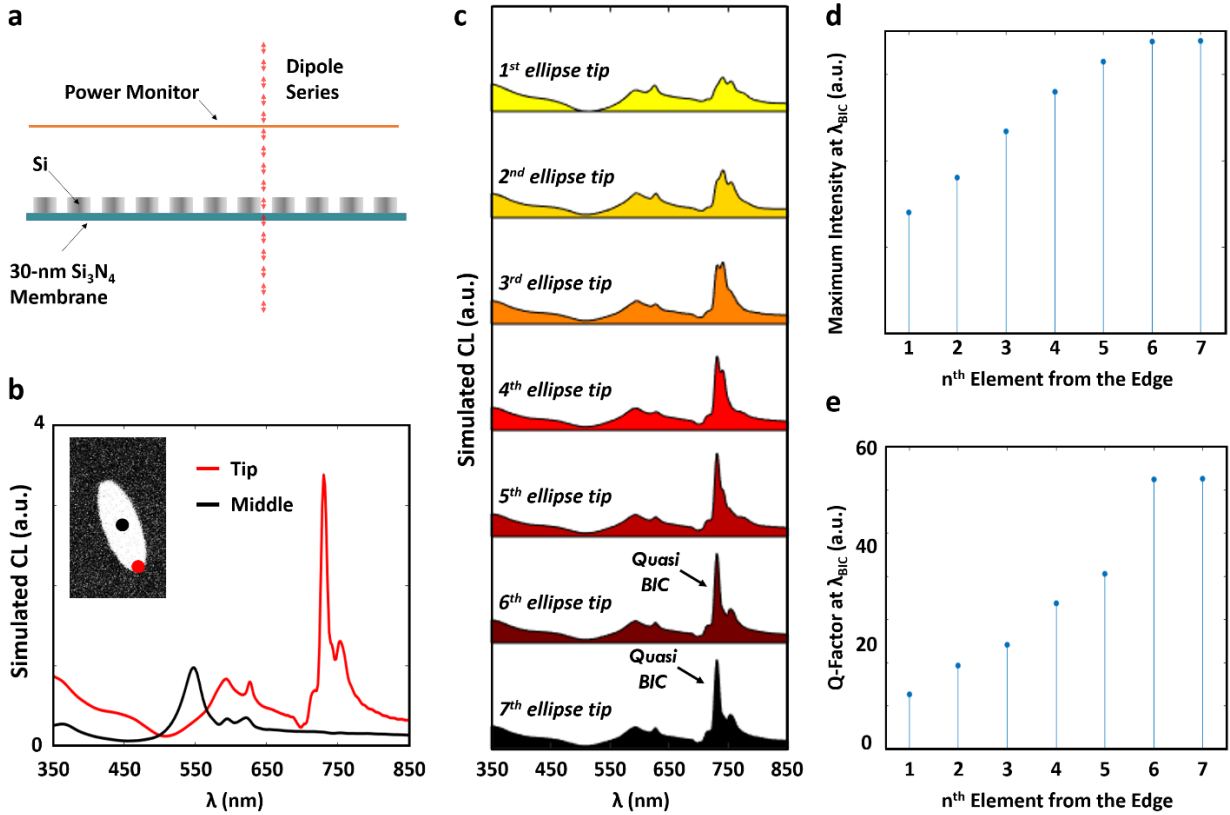

**Supplementary Fig. 4 | Simulation of the CL emission spectra with quasi BIC resonance. a**, Schematic of the $z$-polarized dipole array for simulating the high energy electron beam that excites the CL emission. **b**, Simulated CL spectra when the dipole array is located at the "tip" position and "middle" position of the nanoantennas with a tilt angle $\theta$ of 70°. The simulated CL emission from the "tip" position has a peak around ~720 nm due to the quasi-BIC resonance. **c**, Simulated CL spectrum evolution at the "tip" position of a nanoantenna, from the array edge to the array center. The array has 13×13 elements. It shows that the quasi-BIC resonance is confined within the center of the array, in agreement with the experimental observation shown in Fig. 4d. **d-e**, Evolution of the maximum CL intensity and $Q$-factor at $\lambda_{BIC}$ for the $n^{th}$ ellipse element from the edge.

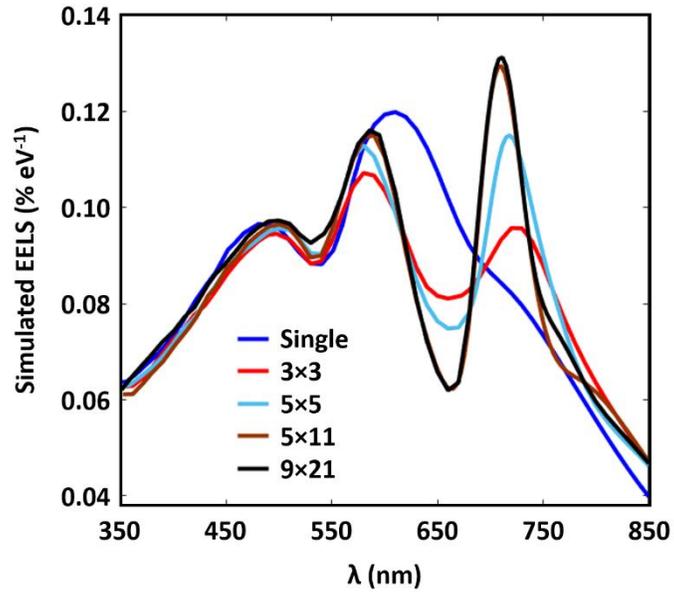

**Supplementary Fig. 5 | Simulated EEL spectra from nanoellipse arrays of different dimensions with θ=90º.** The array dimensions, N×M in the legend indicates the number of elements along the directions defined by the Si ellipse's long (N) and short (M) nanoantenna axes, respectively. It shows how the (Quasi) BIC feature in the spectrum builds up as the number of elements in the array increases.

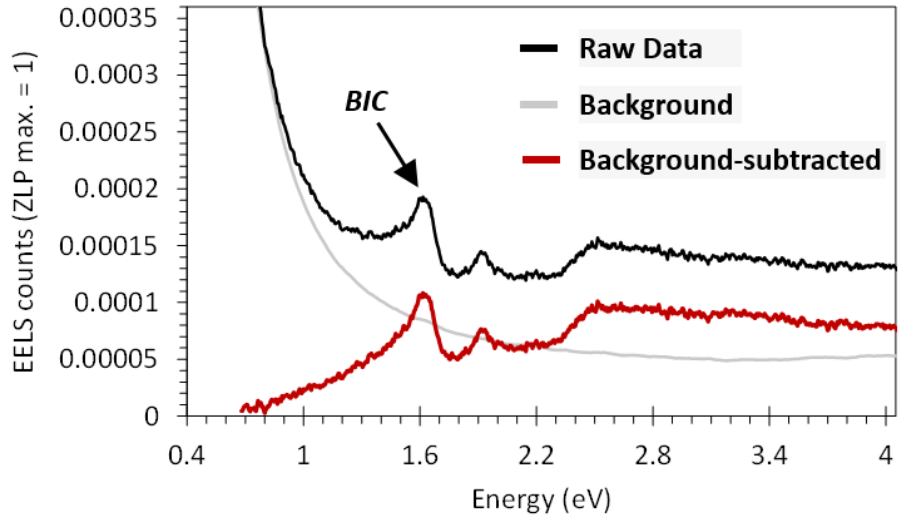

**Supplementary Fig. 6 | Background subtraction of 'zero-loss peak' in the measured EELS spectrum for the nanoantenna array with θ=90°.** The black line presents the raw data of the measured EELS spectrum for the nanoantenna array with θ=90°, which includes the background signal due to the tail of the 'zero-loss peak'. This background signal (grey line) is subtracted by fitting a high-quality zero-loss peak spectrum to the measured spectra between 0.5 and 1.0 eV, well before the first onset of the experimental peaks. After background-subtraction, the remaining signal (red line) is used for the analysis. The high-quality background spectrum was measured from a sample area a few tens of micrometers away from the silicon nanoantennas, with only the 30 nm amorphous $Si_3N_4$ support film present.